\documentclass[reprint,amsmath,amssymb,
 aps,twocolumn,nofootinbib,superscriptaddress]{revtex4-2}

\usepackage{graphicx,braket}
\usepackage{dcolumn}
\usepackage{bm}
\usepackage{hyperref}
\usepackage{cleveref}

\usepackage[dvipsnames]{xcolor}

\crefname{equation}{Eq.}{Eqs.}
\Crefname{equation}{Equation}{Equations}
\crefname{figure}{Fig.}{Figs.}
\Crefname{figure}{Figure}{Figures}
\crefname{section}{Sec.}{Secs.}
\crefname{subsection}{Subsec.}{Subsecs.}
\Crefname{section}{Section}{Sections}
\crefname{appendix}{Appendix}{Appendices}
\Crefname{appendix}{Appendix}{Appendices}

\begin{document}

\title[Article Title]{Passive superconducting circulator on a chip}

\author{Rohit Navarathna}
\email[]{r.navarathna@uq.edu.au}
\affiliation{ARC Centre for Engineered Quantum System, School of Mathematics and Physics,
University of Queensland, Brisbane, QLD 4072, Australia}
\author{Dat Thanh Le} 
\affiliation{ARC Centre for Engineered Quantum System, School of Mathematics and Physics,
University of Queensland, Brisbane, QLD 4072, Australia}
\author{Andr\'es Rosario Hamann}\email{Current address: Department of Physics, ETH Zürich, CH-8093 Zürich, Switzerland}
\affiliation{ARC Centre for Engineered Quantum System, School of Mathematics and Physics,
University of Queensland, Brisbane, QLD 4072, Australia}

\author{Hien Duy Nguyen}
\affiliation{School of Mathematics and Physics, University of Queensland, Brisbane, QLD 4072, Australia}
\author{Thomas M. Stace} 
\affiliation{ARC Centre for Engineered Quantum System, School of Mathematics and Physics,
University of Queensland, Brisbane, QLD 4072, Australia}
\author{Arkady Fedorov}
\email[]{a.fedorov@uq.edu.au}
\affiliation{ARC Centre for Engineered Quantum System, School of Mathematics and Physics,
University of Queensland, Brisbane, QLD 4072, Australia}

\begin{abstract}

An on-chip microwave circulator that is compatible with superconducting devices is a key element for scale-up of superconducting circuits. Previous approaches to integrating circulators on chip involve either external driving that requires extra microwave lines or a strong magnetic field that would compromise superconductivity. Here we report the first proof-of-principle realisation of a passive on-chip circulator which is made from a superconducting loop interrupted by three notionally-identical Josephson junctions and is tuned with only DC control fields. Our experimental results shows evidence for nonreciprocal scattering, and excellent agreement with theoretical simulations. We also present a detailed analysis of quasiparticle 
tunneling in our device using a hidden Markov model. 
By reducing the junction asymmetry and utilising the known methods of protection from quasiparticles, we anticipate that Josephson-loop circulator will become ubiquitous in superconducting circuits. 


\end{abstract}

\keywords{superconducting, passive, circulator, microwave}



\maketitle


\section{Introduction}\label{sec1}

Circulators, nonreciprocal devices that route signals and isolate devices from noise, are a crucial component in cryogenic environments such as superconducting circuits \cite{Pozar11,Gu17}. When used as isolators, they typically route the low power signals from a quantum system to an amplifier, while redirecting noise from the amplifier to a matched load, thus protecting the coherence of the quantum system \cite{metelmannNonreciprocalPhotonTransmission2015,Ruesink16}. Circulators can also function as duplexers that separate the input and output signals to and from a device \cite{Kord18}. This is particularly useful for reflection measurements, where the outgoing signal needs to be separated from the incoming one.

Most commercial microwave circulators harness ferrite components and the Faraday effect to achieve nonreciprocity \cite{Fleury14,Pozar11}, which presents bottlenecks for future development of large-scale superconducting quantum processors.  
First, they are microwave-interference devices, so their size is set by the microwave wavelength. 
Second, the ferrite materials make it impractical to integrate them on a nano-fabricated microwave chip. Furthermore, losses at the cable interconnections for cascaded circulators reduce the quantum efficiency of measurement and control at the quantum noise limit.

Various approaches to miniaturising circulators have been proposed. Some involve driven elements to induce symmetry-breaking fields that require active microwave control lines and consume additional energy \cite{chapmanWidelyTunableOnChip2017, kamalNoiselessNonreciprocityParametric2011, kamalMinimalModelsNonreciprocal2017, estepMagneticfreeNonreciprocityIsolation2014, sliwaReconfigurableJosephsonCirculator2015, lecocqNonreciprocalMicrowaveSignal2017, fangGeneralizedNonreciprocityOptomechanical2017, metelmannNonreciprocalPhotonTransmission2015, petersonStrongNonreciprocityModulated2019, kerckhoffOnChipSuperconductingMicrowave2015, roushanChiralGroundstateCurrents2017, rosenthalBreakingLorentzReciprocity2017}. Others nonreciprocal devices exploit the quantum Hall effect in a two-dimensional gas that necessitates very large magnetic fields \cite{PhysRevLett.93.126804,violaHallEffectGyrators2014, mahoneyOnChipMicrowaveQuantum2017} which are detrimental to superconducting circuits. In this paper, we present the first implementation of an on-chip circulator which is both passive and compatible with superconducting circuits. 
Our design is based on a superconducting ring interrupted by three Josephson junctions, which forms three superconducting islands that are capacitively coupled to external input/output ports \cite{kochTimereversalsymmetryBreakingCircuitQEDbased2010, mullerPassiveOnChipSuperconducting2018}. The device is tuned by only DC control fields that include three charge biases to control the charge distribution on the superconducting islands and a small magnetic flux to break time-reversal symmetry. 
Interference of transitions between the ground state and the excited states of the superconducting ring results in nonreciprocal effects and signal circulation. 

We performed spectroscopic measurements on our device that  confirm theoretical predictions from Ref.\ \cite{mullerPassiveOnChipSuperconducting2018} and allow us to extract the device parameters. 
We also observed quasiparticle tunneling which gives rise to four quasiparticle sectors, each with their own charge configuration and spectroscopic response \cite{leOperatingPassiveOnchip2021}.  

With three readout chains in the dilution refrigerator, we were able to perform three-port device characterisation. Based on our experimental observations, we show that our device exhibits weak nonreciprocity.  However, the excellent quantitative and qualitative agreement with theoretical simulations points to systematic improvements that will produce high fidelity circulation. 
Furthermore, using fast three-port measurements and hidden Markov model analysis \cite{Cappe05,Martinez20,Dixit21} we deduced the tunnelling rates for each of the quasiparticle sectors and determined their average lifetime. We additionally found the presence of a single charge fluctuator as well as charge drift at slower time scales.

\section{
Device description}

\begin{figure}[th!]
    \centering
    \includegraphics[width=0.48\textwidth]{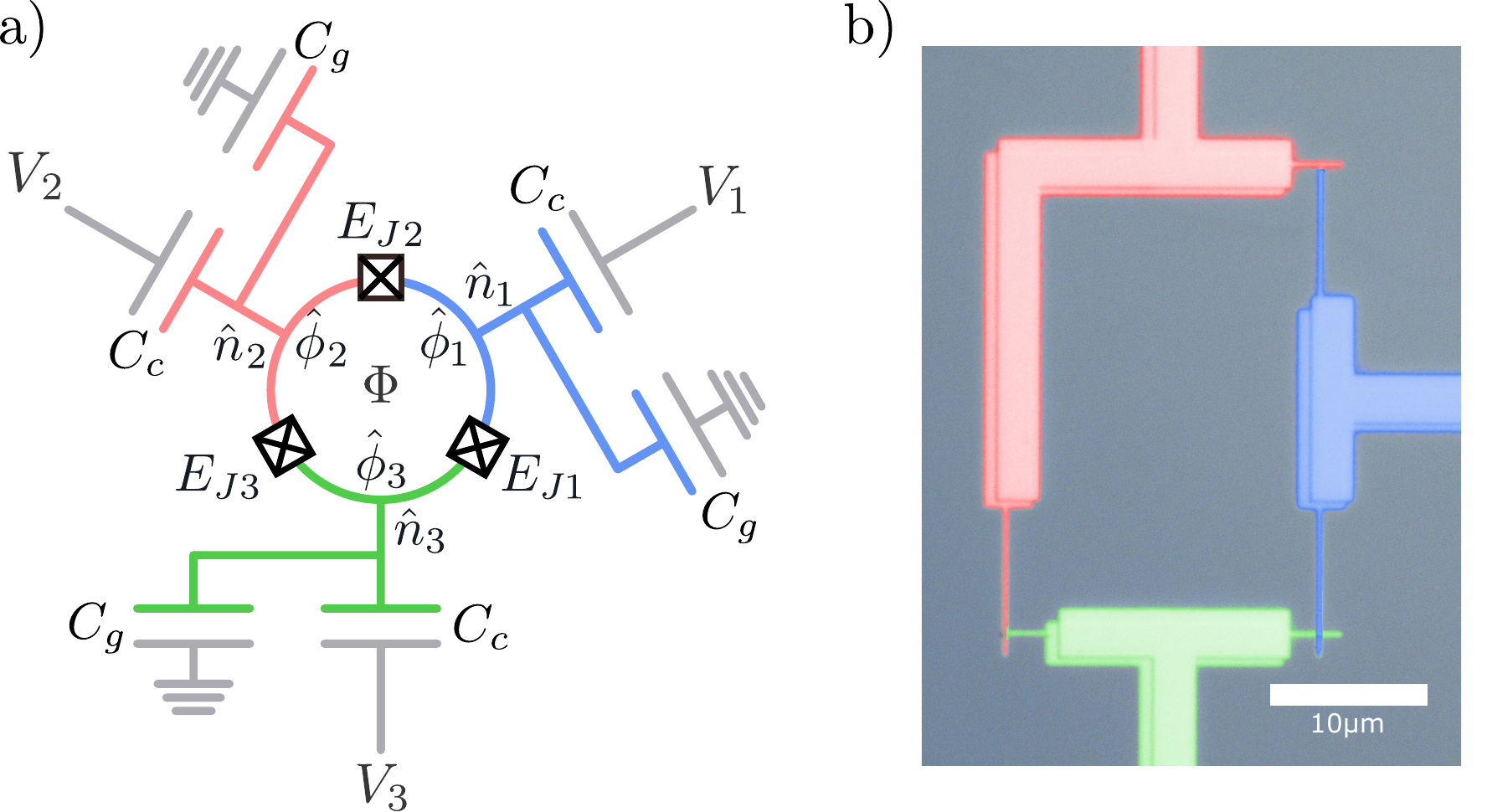}
    \caption{a) Lumped-element circuit of the passive on-chip superconducting circulator. The device is a superconducting loop that is interrupted by three Josephson junctions to form three superconducting islands. The islands are biased by three gate charges and coupled to the external ports by interdigitated capacitors. The superconducting loop is threaded by an external flux.   
    b) False coloured optical microscope image of a fabricated loop. The three islands are red, blue and green. The Josephson junctions are formed by overlapping two layers of aluminium with a layer of aluminium oxide in between.
    }
    \label{fig:circuit}
\end{figure}


Our circulator is schematically displayed as a lumped-element circuit in \cref{fig:circuit}a, consisting of a superconducting loop interrupted by three Josephson junctions. This creates three superconducting islands that are capacitively coupled to  external $50~\Omega $ microwave lines. These lines are the input/output ports, carrying microwave signals and supplying DC bias voltages $V_i$ ($i=1,2,3$) to each superconducting island. The superconducting loop is also threaded by an external flux $\Phi$ provided by a coil. The false-colour optical microscope image in \cref{fig:circuit}b shows a fabricated sample consistent with the lumped-element model. The sample is made of aluminum deposited on a high-resistivity silicon substrate. 
It is mounted to the mixing chamber plate of the dilution refrigerator and connected to three readout chains (see \cref{fig:schematic} for details). This facilitates performing simultaneous three-port characterisation.

The device is described by a Hamiltonian given by \cite{mullerPassiveOnChipSuperconducting2018, leOperatingPassiveOnchip2021}
\begin{equation}
\begin{split}
    \hat H = &(2e)^2(\hat{\textbf{n}} - \textbf{n}_g)\mathbb{C}^{-1}(\hat{\textbf{n}} - \textbf{n}_g) \\
    & - \sum_{i=1}^{3}E_{Ji}\cos{(\hat{\phi}_i - \hat{\phi}_{i+1} - \tfrac{1}{3}\phi)},
\end{split} \label{eq:Hamiltonian}
\end{equation}
where $\hat{\textbf{n}} = \left\{ \hat{n}_1, \hat{n}_2, \hat{n}_3 \right\}$ and $\textbf{n}_g = \left\{ n_{g1}, n_{g2}, n_{g3} \right\}$ are respectively the charge operators and the bias charges (normalised to Cooper-pair charge $2e$)  on the three superconducting islands, 
$\mathbb{C}$ is the capacitance matrix, $E_{Ji}$ are the Josephson energies of the three Josephson junctions, $\hat \phi_i$ are the superconducting phases of the islands, and $\phi = 2\pi\Phi/\Phi_0$ is the reduced flux threading through the superconducting loop. In our experiment, $n_{gi}$ are controlled by applying DC voltages $V_i$  
to the coupling ports through  bias tees at room temperature with $n_{gi} = C_c V_i/(2e)$, whereas $\Phi$ is tuned by a current in a superconducting coil in the same chip with the device as shown in \cref{fig:schematic}.

In the above Hamiltonian, there are two primary energy scales: the charging energy $E_C$ (which is implicit in the first term of \cref{eq:Hamiltonian} but appears explicitly after a coordinate transformation \cite{leOperatingPassiveOnchip2021}) and the mean Josephson energy $\bar{E}_J$. 
We chose the values of $E_C$ and $\bar{E}_J$ such that $\bar{E}_J/E_C\sim 2$, which is in between the Cooper-pair-box \cite{Bouchiat98} and transmon regimes \cite{Koch07}. This, as pointed out in Ref.\ \cite{leOperatingPassiveOnchip2021}, helps increase the bandwidth and relax constraints on junction fabrication, while introducing charge noise to the device. We also selected the working frequency 
of the device to be around 6~GHz, which is within the bandwidth of the amplifiers used in our experiment. 

Signal circulation in the device is mediated by transitions between the ground and excited states of the Hamiltonian $\hat H$ in \cref{eq:Hamiltonian}.  
In particular, we find that in the weak-power limit the scattering matrix element $S_{ij}$, defined by the ratio of the outgoing amplitude at port $i$ to the incoming amplitude at port $j$,  is given by \cite{leOperatingPassiveOnchip2021}
\begin{equation}
     S_{ij} =   \delta_{ij}  - \sum_{k=1,2,\dots} \frac{ \Gamma  \langle k | \hat n_j | 0\rangle \langle 0 | \hat n_i | k\rangle }{i\Delta \omega_k + \Gamma_{k}/2} , \label{eq:Selement}
\end{equation}
where $\ket{0}$ and $\ket{k}$ with $k=1,2, \dots$ are the ground state and the excited states of the Hamiltonian in \cref{eq:Hamiltonian}, 
$\Delta \omega_k$ is the detuning of the excited eigenenergy associated with the excited state $\ket{k}$ from the input driving frequency, $\Gamma$ is the coupling strength between the circulator ring and the external ports, and $\Gamma_{k} $ is the decay rate of the excited state $\ket{k}$ to the ground state $\ket{0}$ due to external couplings.

\section{Spectroscopy and device characterisation} \label{sec:spectroscopic_results}

\begin{figure*}[htb!]
    \centering
    \includegraphics[width=\textwidth]{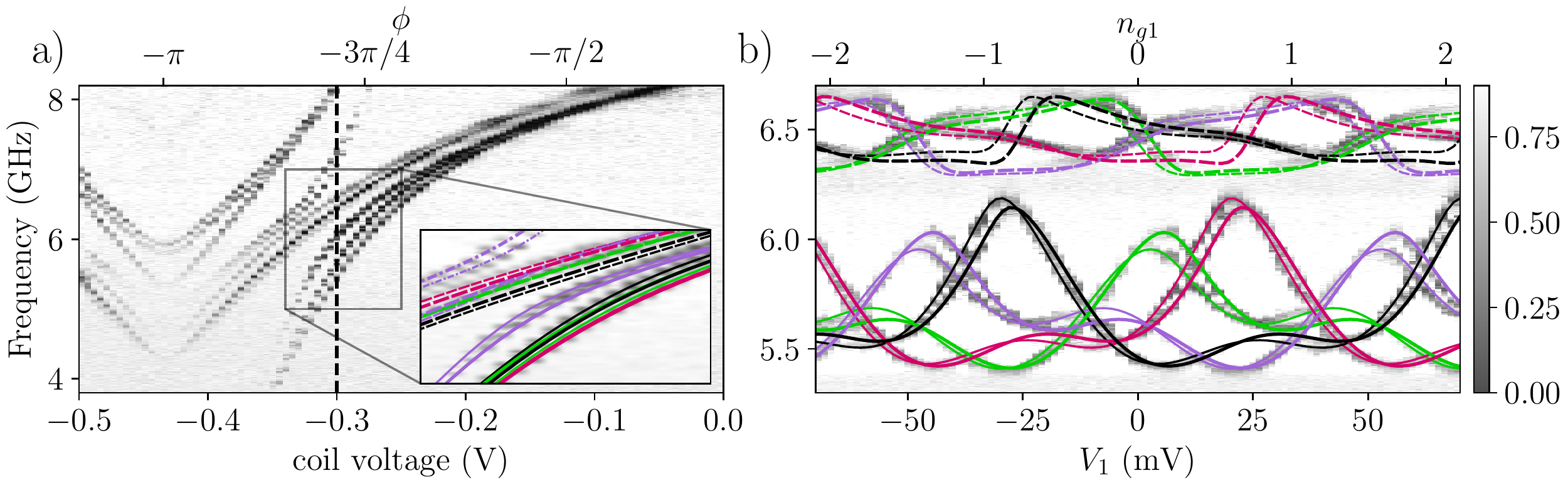}
    \caption{Reflection measurements from port 1 and fit of the device a) to  an external flux ($\phi$) sweep,  b) to a charge bias ($n_{g1}$) sweep at a coil voltage shown by the black dashed line in a). The reflection is measured using a vector network analyzer (VNA) with a bandwidth of 15~kHz, allowing the VNA to capture the drop in reflection before a quasiparticle-tunneling event. This measurement is repeated 100 times, and the minimum for each frequency is plotted as the grayscale density plot after background subtraction. The lines in the inset of a) and in b) are the transition frequencies fitted to the extracted minima for each frequency. There are four transition frequencies (solid, dashed, dash-dotted, dotted) for each quasiparticle configuration (black, purple, green, magenta) and for two charge configurations (two different line thicknesses).  Both fits have the same values of $\{E_C, E_{J1}, E_{J2}, E_{J3} \}/\hbar = \{3.98, 7.85, 8.28, 8.55\}\times \text{GHz}$. We show the full fit for a) in \cref{fig:spectrum_fit_Sup}, .}
    \label{fig:spectrum_fit}
\end{figure*}


We carried out spectroscopic measurements on the device using two ports of a vector network analyzer (VNA). The source and receiver ports of the VNA were respectively connected to one of the input and output microwave cable ports of the device to measure reflection from or transmission through it.


Spectroscopic measurements are shown as a function of the reduced flux $\phi$ in \cref{fig:spectrum_fit}a 
and the charge bias $n_{g1}$ in \cref{fig:spectrum_fit}b. The density plots are the minimum of 100 reflection measurements at a bandwidth of $15$~kHz. This ensures that the transition frequency for each quasiparticle sector was captured in at least one of the 100 measurements. We fit the transition frequencies with the eigenenergies found by solving the eigensystem of the Hamiltonian $\hat H$ in \cref{eq:Hamiltonian}. This allows us to extract the device parameters; in particular, the fit found the charging energy $E_C/\hbar = 3.98$~GHz and the Josephson energies $E_{Ji}/\hbar = \{7.85, 8.28, 8.55\}\times \text{GHz}$, which yields $\bar{E}_J/E_C \sim 2$ and 
matches the values from our room-temperature resistance measurements.
In Ref.~\cite{leOperatingPassiveOnchip2021}, we showed that the tolerable asymmetry of the junctions was 1\% or less, which is determined by the requirement that the junction asymmetry is lower than the external coupling bandwidth (which we estimate to be $\Delta_{\mathrm{BW}}\sim 70$ MHz).  In our device,  the asymmetry of the Josephson energies is much larger than the bandwidth ($\delta E_J\approx 300$ MHz $ \gg \Delta_{\mathrm{BW}}$), 
so the nonreciprocity displayed by the device is limited by the junction asymmetry. 

In \cref{fig:spectrum_fit}a, the transition frequencies are grouped into clusters, each of which contains eight closely located lines. These correspond to two different charge-bias configurations, each  consisting of four charge-parity sectors of the superconducting loop \cite{leOperatingPassiveOnchip2021}  that exchange to each other due to intermittent quasiparticle-tunneling events across the islands. The sectors are represented by the charge parities of two superconducting islands, including $\textsf{e-e}, \textsf{e-o}, \textsf{o-e} $ and $ \textsf{o-o}$, 
 where $\textsf{e}$ ($\textsf{o}$) means an even (odd) number of electrons on a particular island. We note that the superconducting loop is galvanically isolated, so the total charge of the islands is conserved. This makes the charge-parity configuration of the device depend on two out of three islands only, thus justifying the four charge-parity sectors.
 As shown later, we fit a hidden Markov model to stochastic, experimental time-series data to characterise the populations and lifetimes of the quasiparticle sectors. 

In \cref{fig:spectrum_fit}b, we observed two different spectra that belong to the two different charge-bias configurations. We attribute this to the presence of a charge fluctuator located near the superconducting loop. Using our fit routine, we found that the differences in the charge biases of the two configurations are $\{0.11, -0.03, 0.04\}$, indicating that the charge fluctuator is close to the first superconducting island. Moreover, the transition frequencies in \cref{fig:spectrum_fit}b are periodic with $n_{g1}$, as expected for superconducting devices tuned with offset charge biases \cite{Bouchiat98,Koch07}. The periodicity is found to be three Cooper pairs, which is consistent with the fact that there are three superconducting islands in the device. In contrast to the assumption in Ref.~\cite{kochTimereversalsymmetryBreakingCircuitQEDbased2010}, we do not observe discontinuities in the spectrum due to changes in the total charge of the three islands.


Besides the mentioned-above fast charge events, we furthermore observed slow charge drift that occurred roughly every $10$ minutes. This set new values for the charge offsets $\mathbf{n}_g$ and resulted in random change in the transition frequencies. An example of this is shown in \cref{fig:charge_drift_jump}, where we recorded the minimum amplitude of reflection for 100 frequency sweeps.

\section{$S$-matrix characterisation}
Using the four-port VNA and three output readout chains we performed three-port characterisation of our device. Specifically, one port of the VNA was used as the source and connected to a mechanical microwave switch to send signals to one of the three input ports. The three readout chains were connected to the other three ports of the VNA, which allows measuring three S-parameters, $S_{1j}$, $S_{2j}$, and $S_{3j}$, at the same time, where $j$ is the input port. By switching the input port $j$, we obtained the full $S$-matrix.

We note that the scattering matrix measured by the VNA, $\textbf{M}$, is a composite of the scattering matrix from the device, $\textbf{S}$, and the transmission matrices of the input and output lines, which include attenuators and amplifiers, denoted $\textbf{A} = {\rm diag}\{a_1,a_2,a_3\}$ and $\textbf{B} = {\rm diag}\{b_1,b_2,b_3 \}$ respectively. It follows that  ${\bf M=B.S.A}$ where we assumed the external circuitry has perfect impedance matching.

To calibrate the line responses we first changed the flux through the coil such that the frequency of the circulator is off-resonant with the signals from the VNA, and then measured $\textbf{M}$. In this case, we assume that the true $S$-matrix $\textbf{S}$ is lossless and unitary. This gives us the conditions  $\textbf{S} \textbf{S}^\dagger = \mathbb{I}$ and $\textbf{S}^\dagger \textbf{S} = \mathbb{I}$. We measure $\textbf{M}$, and then solve for $\textbf{A}$, $\textbf{B}$, and $\textbf{S}$, which will be close to the identity matrix $\mathbb{I}$ with a small fraction of the signal being transmitted.
We next tuned the circulator frequency to be resonant with that of the VNA signals and used the values of $\textbf{A}$ and $\textbf{B}$ found in the non-resonant case to determine the on-resonance matrix $\textbf{S}$. 

\begin{figure}[t!]
    \centering
    \includegraphics[width=0.48\textwidth]{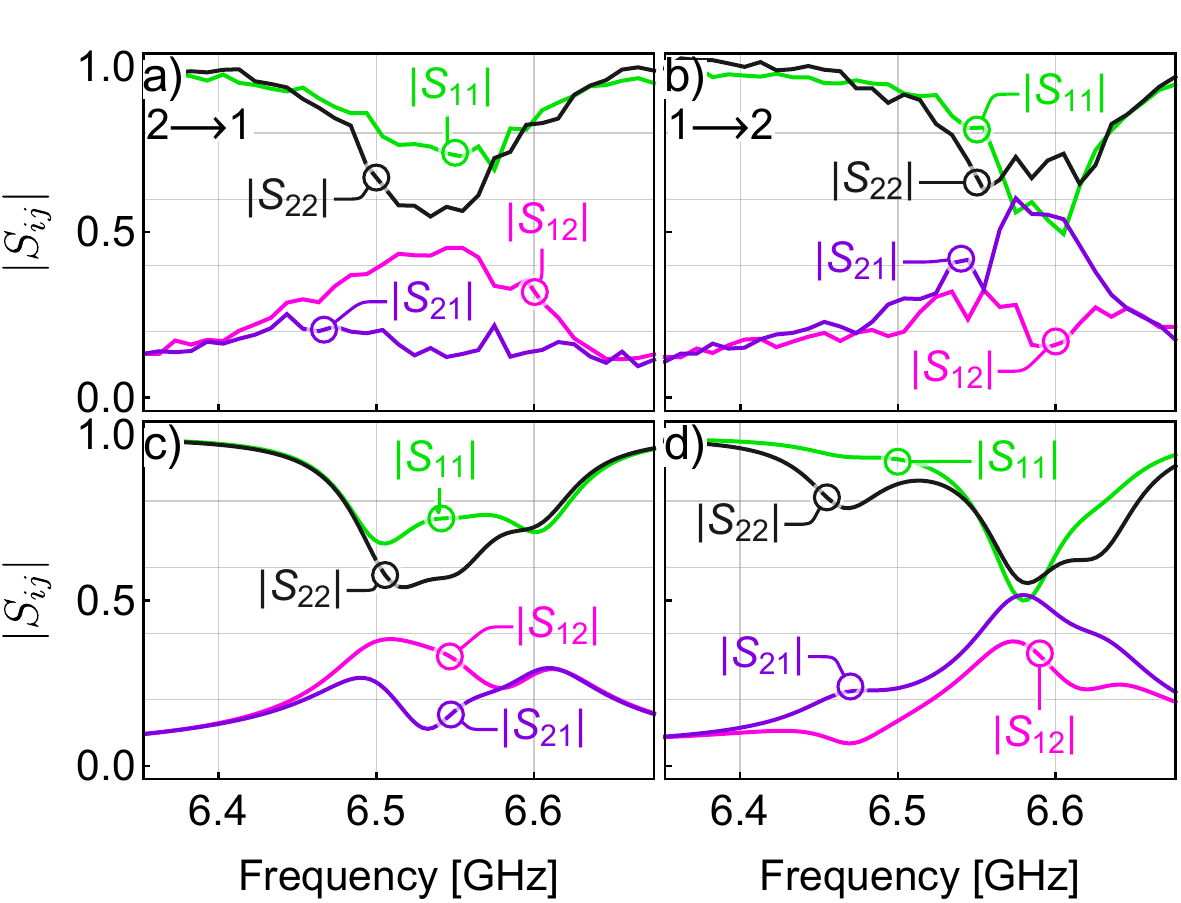}
    \caption{Experimentally extracted partial $S$-matrix elements for two different charge-bias configurations.
    In (a) $|S_{12}|>|S_{21}|$, showing that signals are transmitted more from port 2 to port 1 than from 1 to 2. In (b) a different charge-bias configuration yields $|S_{12}|<|S_{21}|$, reversing the non-reciprocity. 
    The full 3-port $S$-matrix is shown in Fig.~\ref{fig:s_matrix}. (c) and (d) show theoretical simulations corresponding to the experimental configurations above, using the fitting results from \cref{sec:spectroscopic_results} and averaging over the four quasiparticle sectors. We see a good qualitative agreement between experiment and theory.} 
    \label{fig:partial_s}
\end{figure}

Following this procedure, in Fig.~\ref{fig:partial_s} we show a part of the experimentally extracted $S$-matrix, which reveals nonreciprocity in our device. 
In \cref{fig:partial_s}a, we see that when the drive frequency is around 6.5 GHz,  $|S_{12}|>|S_{21}|$. 
At this frequency, the reflections $|S_{11}|$ and $|S_{22}|$ are also at minima. 
 \Cref{fig:partial_s}b shows that by tuning the charge-bias voltages $V_1$, $V_2$ and $V_3$ (at constant flux bias) we can tune the device so that $|S_{12}|<|S_{21}|$; that is, we are able to electronically change the direction of signal transfer.  We provide the full $S$-matrix for both circulation directions in \cref{app:fullS} (see \cref{fig:s_matrix}).

In Figs.\ \ref{fig:partial_s}c and d, we performed simulations to reconstruct the $S$-matrix elements in Figs.\ \ref{fig:partial_s}a and b. To this end, we use the  device parameters measured in \cref{sec:spectroscopic_results} and for each working point (including the external biases and the signal frequency) we compute the $S$-matrix for all four quasiparticle sectors, and then average over the sectors (with equal weights) to achieve an averaged $S$-matrix. 
Comparing Figs.\ \ref{fig:partial_s}a and b to Figs.\ \ref{fig:partial_s}c and d respectively, we found qualitative agreement between theory and experiment. 

\begin{figure}[t!]
    \centering
    \includegraphics[width=0.48\textwidth]{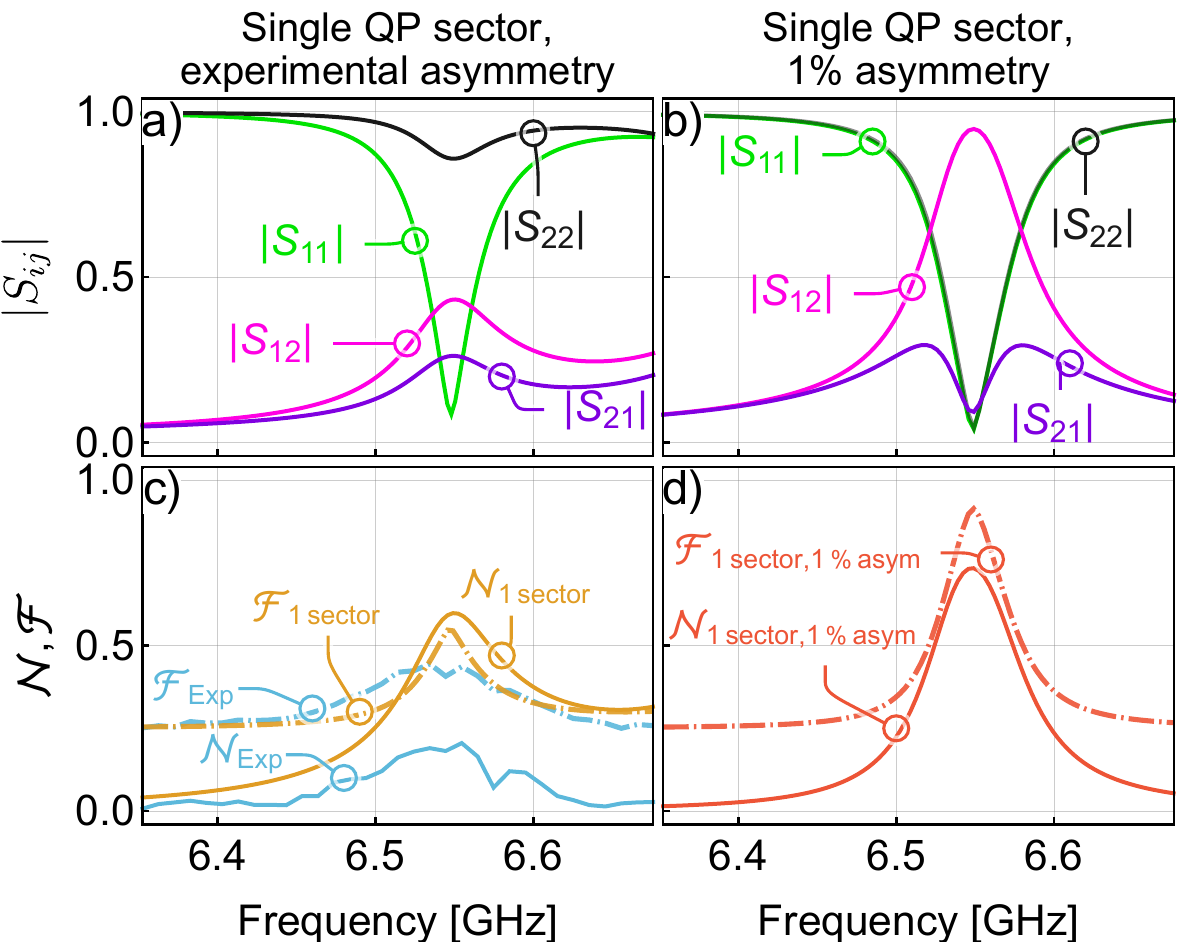}
    \caption{ (a) Simulated $S$-matrix elements when the device stays in a single quasiparticle (QP) sector with the same junction asymmetry as in the current sample. (b) Simulated $S$-matrix with junction asymmetry reduced to 1\%, with $E_{J1}/E_{J2}=0.99$ and $E_{J3}/E_{J2}=1.01$. (c) Nonreciprocity and fidelity  defined in Eqs. \eqref{eq:Nonreciprocity} and \eqref{eq:Fidelity} for the experimentally measured $S$-matrix, and for the simulated $S$-matrix with the experimentally determined values of $E_J$ but no QP tunnelling. 
    (d) Simulated $S$-matrix with no QP tunnelling, and 1\% junction dispersion. }
    \label{fig:partial_s_Idealised}
\end{figure}

Given the deleterious effect of quasiparticles and junction fabrication asymmetry on circulation, we computed the potential performance of the device in two scenarios that remove these two effects. 
First, we computed the performance assuming that the system dwells in a single quasiparticle sector.  The simulated $S$-matrix for this case is in \cref{fig:partial_s_Idealised}a, where we see a much stronger effect on resonance, $S_{11}\ll1$, but only modest changes in the off-diagonal elements.  
Second, we compute the performance when reducing the junction asymmetry, in addition to suppressing quasiparticle tunneling.  In \cref{fig:partial_s_Idealised}b, we show the $S$-matrix elements assuming 1\% junction asymmetry, i.e.\ $E_{J1}/E_{J2}=0.99$  and $E_{J3}/E_{J2}=1.01$.
From these simulations we estimate the bandwidth within a single quasiparticle sector to be around $70$ MHz.

To quantify the observed and simulated circulation, we introduce the \emph{nonreciprocity}, $\mathcal{N}$   \cite{Caloz18}, and the \textit{fidelity}, $\mathcal{F}$ \cite{Scheucher16}
\begin{eqnarray}
\mathcal{N} &=&  || \mathbf{S} - \mathbf{S}^{\intercal}  ||/\sqrt{8}, \label{eq:Nonreciprocity} \\
\mathcal{F} &=& 1 -  {{\sum}_{i,j}} \big | |S_{ij}| - |S^{\mathrm{ideal}}_{ij}| \big|/8, \label{eq:Fidelity}
\end{eqnarray}
where $ \lVert X \rVert = \sqrt{ \mathrm{Tr}(X X^\dag)} $ denotes the norm of a matrix $X$, $\sqrt{8}$ is a normalisation factor, and $S^{\rm ideal}_{ij}$ are the elements of an ideal circulator scattering matrix. We comment more on the nonreciprocity measure in \cref{subsec:Measure_Nonreciprocity}.   
. 


\begin{figure*}[ht!]
    \centering
    \includegraphics[width=\textwidth]{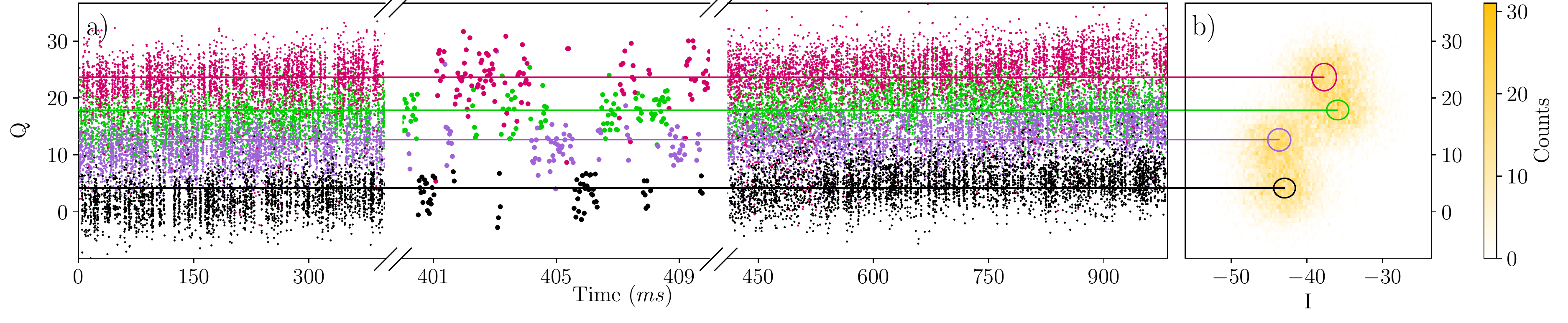}
    \caption{a) Repeated measurement of $S_{11}$ of the device at a frequency of $6.709$~GHz, an external flux $\phi=1.9$. 
    The out-of-phase quadrature Q versus time shows gaussian noise that switches  between one of four unobserved states.  We fit the the full $3\times3$ scattering matrix data, $\bf{S}$,  with a four-state hidden Markov model (HMM), using approximately 32k time samples, and attribute the data to one of the four hidden states accordingly, indicated by color.  We interpret these states as the quasiparticle states seen \cref{fig:spectrum_fit}. b) Histogram of the $S_{11}$ data in the IQ plane, showing clustering of $S_{11}$. 
    We also show the  $1\sigma$ confidence ellipse extracted from the HMM (circles).  We note that the separation in the hidden state data is greater than observed here, which is merely a  projection of  $\bf{S}$ onto a 2-dimensional subspace.  
    }
    \label{fig:time-series}
\end{figure*}

In \cref{fig:partial_s_Idealised}c, we show the nonreciprocity $\mathcal{N}$ for both the experimentally measured $S$-matrix in \cref{fig:partial_s}a, $\mathcal{N}_{\mathrm{exp}}$, and the `no-quasiparticle-tunneling' $S$-matrix in \cref{fig:partial_s_Idealised}a, $\mathcal{N}_{\mathrm{1\,sector}}$.  We see that the nonreciprocity for the latter is considerably improved, while the fidelities of the two cases, $\mathcal{F}_{\mathrm{exp}}$ and $\mathcal{F}_{\mathrm{1\,sector}}$, are comparable, indicating that a higher nonreciprocity is necessary but not sufficient for better circulation. 

Reducing the junction asymmetry to $1\%$, in addition to suppressing quasiparticle tunneling,  greatly boosts the device performance.  \Cref{fig:partial_s_Idealised}d  shows theoretical predictions for the fidelity and nonreciprocity in the absence of quasiparticle fluctuations, and assuming parameter variations are reduced to 1\%.  We see both the nonreciprocity and the fidelity are substantially improved relative to the experimental results.  



\section{Quasiparticle tunnelling}

Due to the significant impact of quasiparticle tunneling, the measurement setup 
was modified to measure the full $S$-matrix faster than the quasiparticle tunnelling rate. 
For this purpose, we used a combination of microwave sources, an AWG, and a digitizer (shown in \cref{app:measurement}, \cref{fig:circ_time_setup}), which allows us to measure the $S$-matrix in $30~\mu$s.  

In order to distinguish the four distinct states corresponding to the four quasiparticle sectors, we use a relatively high power, -100~dBm, at the device to enhance the single-shot signal-to-noise ratio for the scatter matrix measurements.  We took a continuous run of 32,768 time-series measurements of the full scattering matrix, and then employed a hidden Markov model (HMM) to analyse the time-series data (see \cref{sec:hmm} for details) \cite{Cappe05,Martinez20,Dixit21}.  At each time step, the HMM  is able to make a statistical inference about the hidden quasiparticle state from the full $3\times3$ scattering matrix, and the emission statistics from the data within each hidden state. 

In \cref{fig:time-series}a, we show the Q-quadrature of the $S_{11}$ component of the measured $S$-matrix, over time. We observe switching of the scattering matrix, corresponding to jumps between the four  quasiparticle  sectors that were observed in \cref{fig:spectrum_fit}. At each time, we classify the data as being associated to one of the four hidden states, indicated by the colours.  In  \cref{fig:time-series}b, we plot the data in the IQ plane which reveals four distinct clusters, one per quasiparticle sector, along with the standard-error ellipse for each sector extracted from the HMM fitting.

In these measurements, we were also able to extract the average $S$-matrix for each quasiparticle sector. However, because of the high-power signals, our $S$-matrix elements showed only very minor nonreciprocity. With quantum limited amplifiers, one could classify the quasiparticle sectors at a lower power. In this case, we expect to see higher circulation for one of the quasiparticle sectors as theoretically described in Ref.\  \cite{leOperatingPassiveOnchip2021}.

The HMM provides additional information about the the dynamics of the jump processes.  In particular, we find that the jumps between the quasiparticle states are well-described by Poisson processes with decay times of around 200~$\mu$s in each quasiparticle sector (see \cref{sec:hmm}, \cref{fig:jump_rates}).

\section{Conclusions}

In conclusion, we fabricated and characterised a passive superconducting circulator on a chip. Our device is fully compatible with superconducting qubit fabrication and can be tuned with only DC control. Our experimental results showed  clear evidence of nonreciprocity, but not  sufficient for high-fidelity circulation.  

We provide strong evidence that quasiparticle fluctuations and device asymmetry are significant and detrimental to the operation of the device. 
We showed that  the device nonreciprocity and circulation can be substantially improved by suppressing quasiparticle tunnelling such that the device stays in one single quasiparticle sector, and reducing the disparity in the three Josephson junctions energies to $1\%$ or less. 

The precision of our Josephson junction fabrication is comparable to the standard shadow evaporation fabrication. However, post-fabrication laser annealing has been shown to improve the variation of junction normal resistances to be  $\sim 0.3\%$ of design values across a device \cite{hertzbergLaserannealingJosephsonJunctions2021}.  This level of asymmetry is suitable for reaching  the necessary junction precision $\lesssim 1\%$  required for high-quality circulation. 
In addition,  recent progress has been made in using different materials, including normal metals \cite{Mannila21} or low-gap superconductors \cite{martinisSavingSuperconductingQuantum2021a}, to suppress quasiparticle tunneling.  Together with possible reduction of charge sensitivity, these provide a pathway to enhance the performance of the device we have demonstrated towards practically useful components for on-chip superconducting quantum processors.

\begin{acknowledgments}
We thank Clemens Müller and Prasanna Pakkiam for useful discussions.
This research was supported by the Australian Research Council Centres of Excellence for Engineered Quantum Systems (Projects No.\ CE170100009 and No.\ CE110001013).   
\end{acknowledgments}

\bibliography{paper}

\clearpage
\appendix

\section{Fabrication and experiment setup}

\begin{figure}[ht!]
    \centering
    \includegraphics[width=0.49\textwidth]{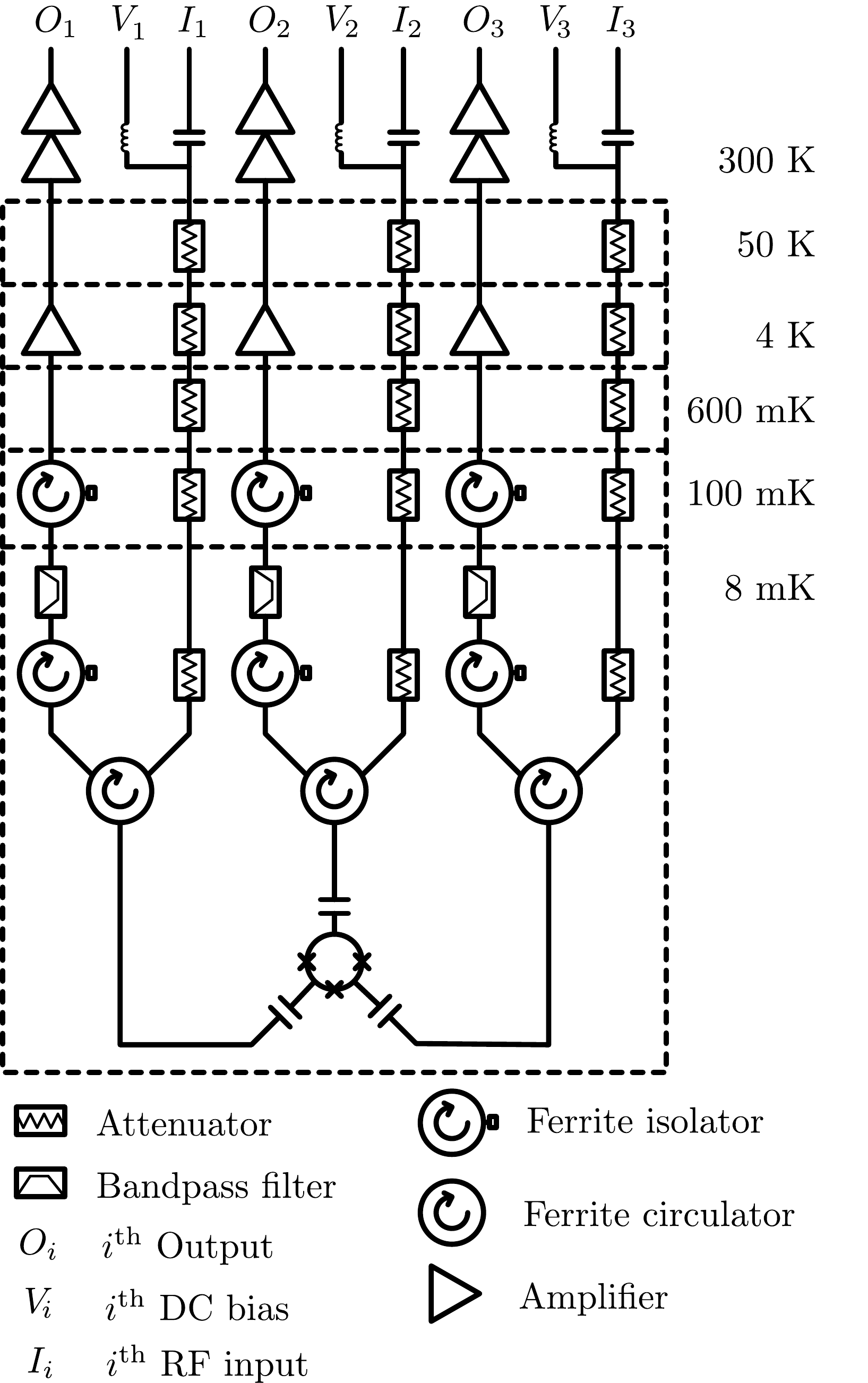}
    \caption{Schematic of the setup inside the dilution refrigerator. Each of the three ports of the device is connected to a ferrite circulator to both send and receive signals from the device. The inputs $I_i$ are sent through attenuators to the device. A DC voltage $V_i$ is added to the RF input via bias tees at room temperature. Each output $O_i$ from the device passes through a circulator, two isolators, a band pass filter, a HEMT amplifier and room temperature amplifiers.}
    \label{fig:schematic}
\end{figure}

The device was fabricated on a $500~\mu$m thick high-resistivity silicon substrate. $300~\mu$m of the substrate was pre-diced on the back side to allow cleaving of the chip rather than dicing. This is because dicing requires an extra layer of spin-coating, which would increase the spread in $E_{Ji}$ of the Josephson junctions. A bi-layer PMGI-PMMA resist stack was spin-coated on the high resistivity silicon substrate before patterning using electron beam lithography with a Raith EBPG. The patterned chip was developed and the native silicon oxide was removed with a HF dip prior to loading the chip into a Plassys MEB550 electron beam evaporator. In order to make the Josephson junctions, two layers of aluminium were evaporated on the substrate which was tilted at a 45 degrees to the vertical. The first and second layer of thicknesses 20 nm and 60 nm were evaporated by rotating the substrate holder along the normal of the substrate by 0 degrees and 90 degrees (planetary rotation). This is done to make a quasiparticle trap via engineering the superconducting gap \cite{martinisSavingSuperconductingQuantum2021a}. The thinner layer of aluminium has a superconducting gap higher than the rest of the aluminium, which acts as a potential barrier for quasiparticle tunnelling. We still observe quasiparticle tunnelling with the quasiparticle trap. However, we could not measure the reduction in the tunnelling rate as we do not have a reference tunnelling rate for a sample without a trap. After evaporation, the chip was cleaved along the pre-diced lines using a cleaver and carefully placed tweezers. The resist was then removed using PG remover heated to $70^\circ$~C. The room temperature resistance of the Josephson junctions were determined using a probe station. The $E_{Ji}$ were approximated to be 7.90 GHz, 8.10 GHz and 8.36 GHz, corresponding to a separation of $5.6\%$.

The schematic of our experiment is shown in \cref{fig:schematic}. The sample was loaded in the base plate of a Bluefors LD dilution refrigerator and cooled down to $8$~mK. The three ports of the circulator were connected to input and output microwave cables through three commercial ferrite circulators to measure transmission and reflection from all three ports 
The flux through the loop $\phi$ and the charge configuration $\textbf{n}_g$ is controlled by a DC voltage source connected to a coil and three bias tees at room temperature respectively.

\section{Fitting and  simulated scattering matrix elements} \label{subsec:theory}

\begin{figure}[!htbp]
    \centering
    \includegraphics[width=0.48\textwidth]{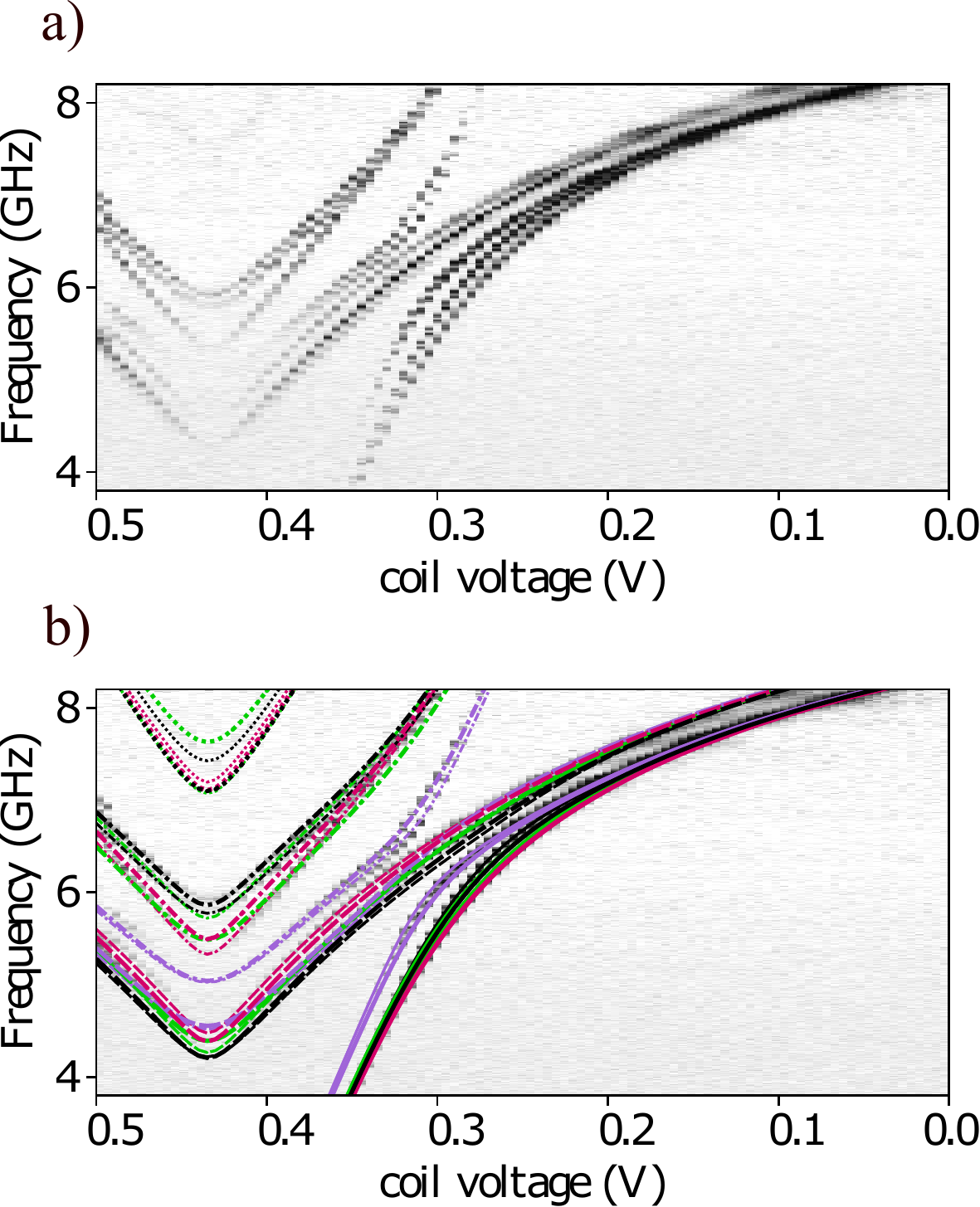}
    \caption{ Same as \cref{fig:spectrum_fit}a but a) raw data, and b) including the full fits. 
    The fitted lines are four transition frequencies (solid, dashed, dash-dotted, dotted) for each quasiparticle configuration (black, purple, green, magenta) and for two charge-bias configurations (two different line thicknesses). }
    \label{fig:spectrum_fit_Sup}
\end{figure}

The theoretical fit in \cref{fig:spectrum_fit} consists of the transitions from the ground state to the excited states of the Hamiltonian $\hat H$ in \cref{eq:Hamiltonian}. We find these transitions by solving the eigenenergies of $\hat H$. To do this, we first perform a coordinate transformation
\begin{eqnarray}
& \hat n'_1 = \hat n_1, \hspace{0.25cm} \hat n'_2 = - \hat n_2, \hspace{0.25cm} \hat n'_3 = \hat n_1 + \hat n_2 + \hat n_3 = n_0, \label{eq:CoorTransform} \\
& \hat \phi'_1 = \hat \phi_1 - \hat \phi_3, \hspace{0.25cm} \hat \phi'_2 = \hat \phi_3 - \hat \phi_2, \hspace{0.25cm} \hat \phi'_3 = \hat \phi_3,
\end{eqnarray}
where $n_0$ is the conserved total charge number, which is set to be zero in our simulations. In the new coordinates, the Hamiltonian  is \cite{leOperatingPassiveOnchip2021}
\begin{eqnarray}
    && \hat H' (n_{g1}, n_{g2}, n_{g3})  \nonumber \\
   &=& E_{C} \big( (\hat n'_1 - \tfrac{1}{2}(n_0 +n_{g_1}- n_{g_3}))^2 \nonumber  \\
&& + (\hat n'_2 + \tfrac{1}{2}(n_0 +n_{g_2}- n_{g_3}))^2 -\hat n'_1 \hat n'_2 \big) \nonumber \\
&& -E_{J1}  \cos(\hat \phi'_1 - \tfrac{1}{3}\phi) - E_{J2}\cos(\hat \phi'_2- \tfrac{1}{3}\phi) \nonumber \\
&& - E_{J3} \cos(\hat \phi'_1 + \hat \phi'_2  + \tfrac{1}{3}\phi)  \label{eq:HamiltonianNewCor},
\end{eqnarray}
where $E_{C}$ is the charging energy. We then numerically diagonalise $\hat H'$ in the truncated discrete charge basis. 

The Hamiltonian $\hat H'(n_{g1},n_{g2},n_{g3})$ in Eq.~\eqref{eq:HamiltonianNewCor} describes one of the four quasiparticle sectors $\textsf{ee}, \textsf{eo}, \textsf{oe}$ and $\textsf{oo}$. The Hamiltonians of the other sectors are given by $\hat H'(n_{g1}-\tfrac{1}{2},n_{g2}+ \tfrac{1}{2},n_{g3})$, $\hat H'(n_{g1},n_{g2}-\tfrac{1}{2},n_{g3}+\tfrac{1}{2})$, and $\hat H'(n_{g1}+\tfrac{1}{2},n_{g2},n_{g3}-\tfrac{1}{2})$. Four more Hamiltonians are used to account for the presence of a charge fluctuator in \cref{fig:spectrum_fit}b, for which small fluctuations $\delta n_{gi}$ were added to $n_{gi}$. In \cref{fig:spectrum_fit_Sup},  we show the full fit of \cref{fig:spectrum_fit}a, where one can see an excellent overlap between the experimental data and the fitted transitions for the whole range of the sweep.

The simulated scattering matrix elements in Figs.~\ref{fig:partial_s}c and d and Figs.~\ref{fig:partial_s_Idealised}a and b are computed using $S_{ij}$ given in \cref{eq:Selement}. Within this expression,  the charge matrix elements $\langle k| \hat n_j | 0 \rangle$ and $\langle 0 | \hat n_i |k \rangle$, the detunings $\Delta \omega_k$, and the decay rates $\Gamma_k$ are found from the eigensystem of the Hamiltonian $\hat H'$ for each quasiparticle sector \cite{leOperatingPassiveOnchip2021}. The coupling strength $\Gamma$ can be varied in our simulation to fit to the bandwidth from the experimental data. We also note that $S_{ij}$ in \cref{eq:Selement} is an approximate expression for the $S$-matrix elements that is obtained from adiabatic elimination of the open waveguide-ring system \cite{leOperatingPassiveOnchip2021}. However, we found that  using $S_{ij}$ in \cref{eq:Selement} or $S_{ij}$ computed from solving an open-system SLH master equation \cite{mullerPassiveOnChipSuperconducting2018} to simulate the $S$-matrix yields essentially the same results.

\section{Charge drifts and jumps}

\begin{figure*}[!htbp]
    \centering
    \includegraphics[width=\textwidth]{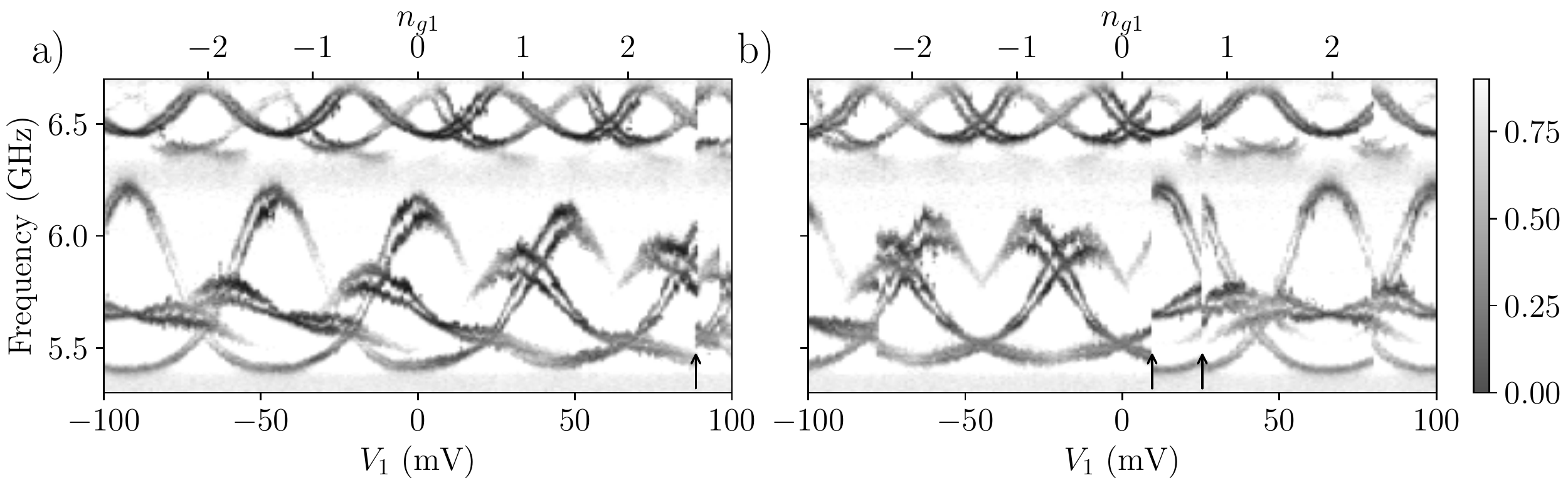}
    \caption{a) and b) Two examples of reflection measurements where $n_{g1}$ is swept with an external voltage using the same method as in Fig.~\ref{fig:spectrum_fit}. 
    Both the plots are measured from left to right in time. We can see charge reconfiguration events shown by the black arrows, where the transition frequencies witness a discontinuous change.}
    \label{fig:charge_drift_jump}
\end{figure*}

Figure \ref{fig:charge_drift_jump} shows reflection measurements where $n_{g1}$ is swept with an external voltage using the same method as in Fig.~\ref{fig:spectrum_fit}. The plots were recorded in time from left to right. Charge jumps  manifest via sudden changes in the transition frequencies, which are marked by black arrows in \cref{fig:charge_drift_jump}. Charge drift can be observed as a gradual change in the periodic pattern in \cref{fig:charge_drift_jump}a.

\section{Full experimentally extracted scattering matrix}\label{app:fullS}

\begin{figure}[!htbp]
    \centering
    \includegraphics[width=0.48\textwidth]{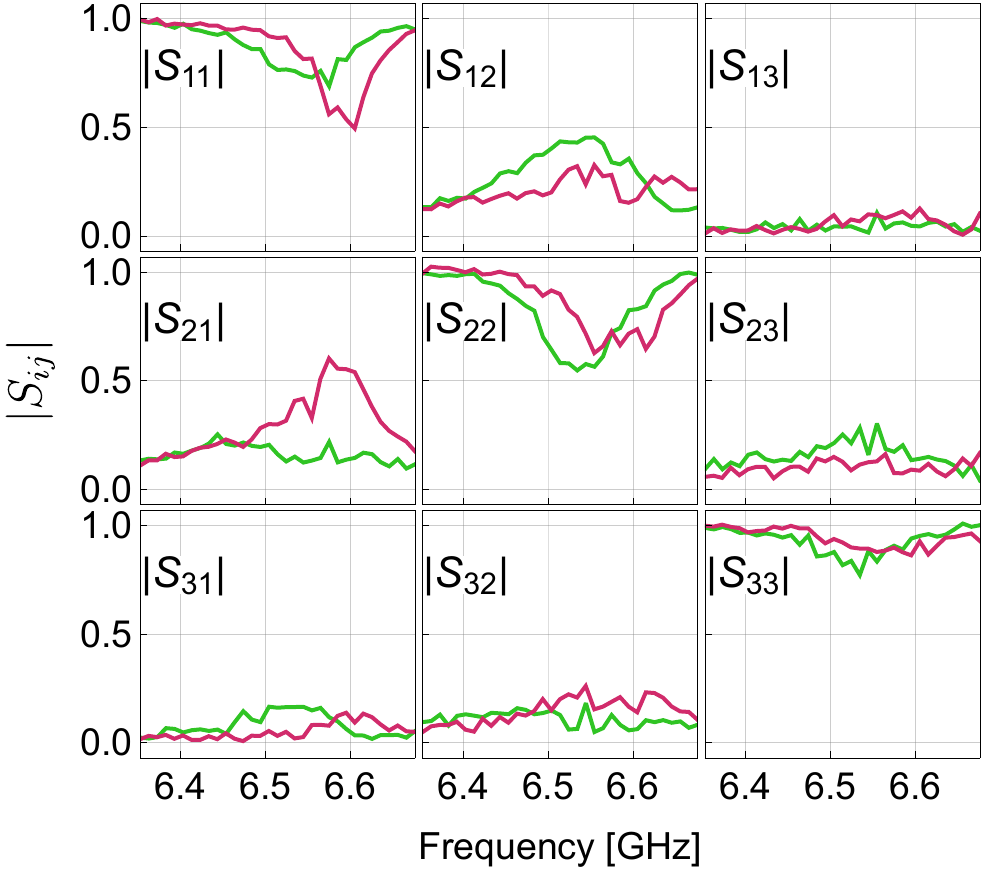}
    \caption{Full scattering matrices of a) the partial one in \cref{fig:partial_s}a and b) the partial one in \cref{fig:partial_s}b.}
    \label{fig:s_matrix}
\end{figure}

Figure \ref{fig:s_matrix} displays the full experimentally extracted scattering matrix for both circulation directions in Figs.~\ref{fig:partial_s}a and b. For both cases,  nonreciprocity exhibits clearly when  comparing $S_{12}$ and $S_{21}$, but not with the pair $S_{13}$ and $S_{31}$ or $S_{23}$ and $S_{32}$. Signal transmissions as seen from the figure are also low.   

\section{Measures of nonreciprocity and circulation} \label{subsec:Measure_Nonreciprocity}

The measure of nonreciprocity $\mathcal{N}$ defined in Eq.~\eqref{eq:Nonreciprocity} is too broad for capturing circulation, as seen by considering a 3-port gyrator which has the following scattering matrix
\begin{equation}
\mathbf{S}_{\rm gyr}=\left[\begin{array}{ccc}
0 & 1 & 0 \\
-1 & 0 & 0 \\
0 & 0 & 1\end{array}\right]. 
\end{equation}
For this matrix, $\mathcal{N}(\mathbf{S}_{\rm gyr})=1$, however it is a poor-quality circulator (we note here that $\lVert \mathbf{S}_{\rm gyr} - \mathbf{S}_{\rm gyr}^{\intercal} \rVert = \sqrt{8}$, explaining the normalisation factor in Eq.~\eqref{eq:Nonreciprocity}). We therefore introduce the fidelity $\mathcal{F}$ defined in Eq.~\eqref{eq:Fidelity}.  
For an ideal circulator with 
\begin{equation}
\mathbf{S}^{\rm ideal}=\left[\begin{array}{ccc}
0 & 1 & 0 \\
0 & 0 & 1 \\
1 & 0 & 0\end{array}\right] \text{or} \left[\begin{array}{ccc}
0 & 0 & 1 \\
1 & 0 & 0 \\
0 & 1 & 0\end{array}\right],
\end{equation}
we find $\mathcal{N}(\mathbf{S}^{\rm ideal}) = \sqrt{3/4}$ and $\mathcal{F} (\mathbf{S}^{\rm ideal}) = 1$.

\section{Time-series measurement setup}\label{app:measurement}

\begin{figure*}[!htbp]
    \centering
    \includegraphics[width=\textwidth]{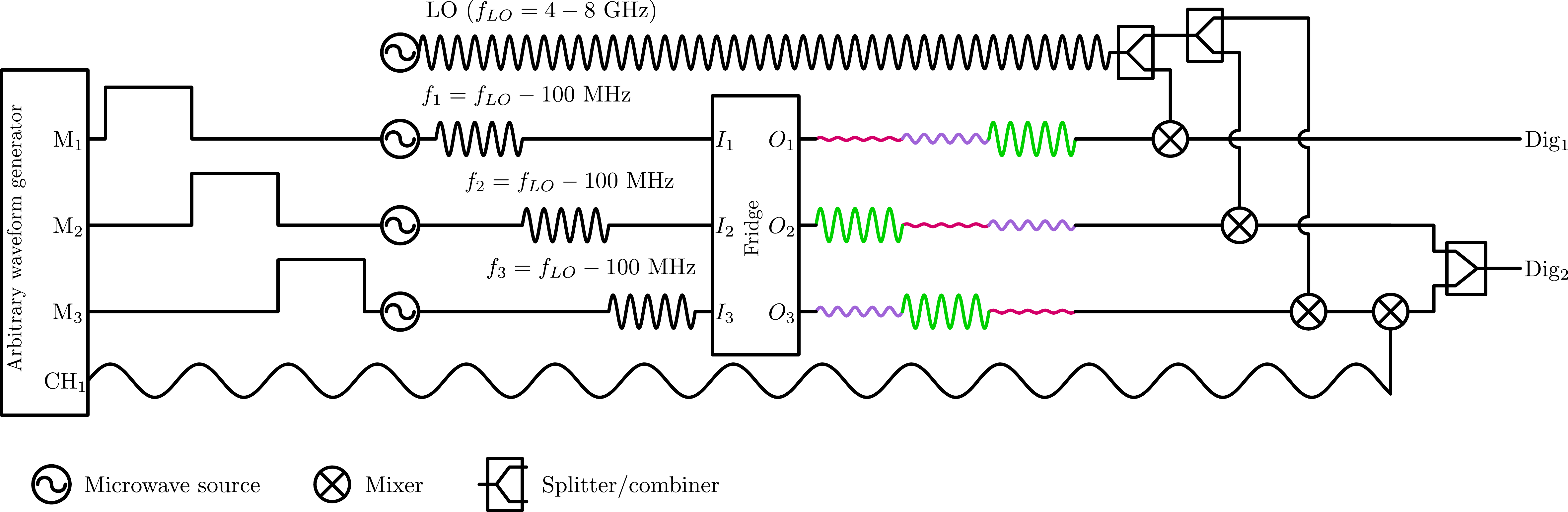}
    \caption{Schematic of the measurement hardware setup for the time series measurements. The marker channels of the AWG are used to send digital signals that gate the output of three microwave sources. The same measurement frequency $f$ is the output from the three sources, one after the other in time. The $S$-matrix is imprinted in the three outputs. The colors of the outputs refer to the ideal clockwise circulation (green), ideal counter-clockwise circulation (purple), and reflection (magenta). A fourth microwave sources is used as an LO with a frequency of $f - 100$~MHz for all three outputs using splitters. The third output is further separated in frequency using another mixer and a continuous sine wave used as the LO from an AWG channel. The second and third outputs are then combined. The first output, along with the combined output is recorded on the digitizer.}
    \label{fig:circ_time_setup}
\end{figure*}

Figure \ref{fig:circ_time_setup} is a schematic diagram of our time-series measurement setup, which uses four microwave sources, an AWG, and a digitizer. This facilitates fast measurements of the $S$-matrix at a rate of $30~\mu\rm s$ to capture the dynamics of quasiparticle events.

\section{Hidden Markov model}\label{sec:hmm}
A hidden Markov model (HMM) \cite{Cappe05,Martinez20,Dixit21} is a statistical tool that can be used to predict the most likely states of a system given a sequence of observed data. A HMM consists of a transition matrix that contains the transition probabilities between hidden state, and the emission process that describes the emitted data within each hidden state.  If the process model is not known beforehand, the Baum-Welch algorithm can be used to  fit a model to the observed data. 

In our case, the time series of the full $3\times3$ $S$-matrices are the observed data, and the four quasiparticle configurations are the hidden states. We used the \textit{pomegranate} Python package \cite{schreiber2018pomegranate} to fit the observed data.  \textit{Pomegranate}  first finds clusters of data using the k-means algorithm and then uses the Baum-Welch algorithm to extract the HMM. With the fitted HMM, we can then predict the most likely sequence of quasiparticle configurations that led to the observed data. This predicted sequence of quasiparticle configurations were used to classify the data in Fig.~\ref{fig:time-series}a.

\section{Quasiparticle jump statistics}

\begin{figure*}
    \centering
    \includegraphics[width=\textwidth]{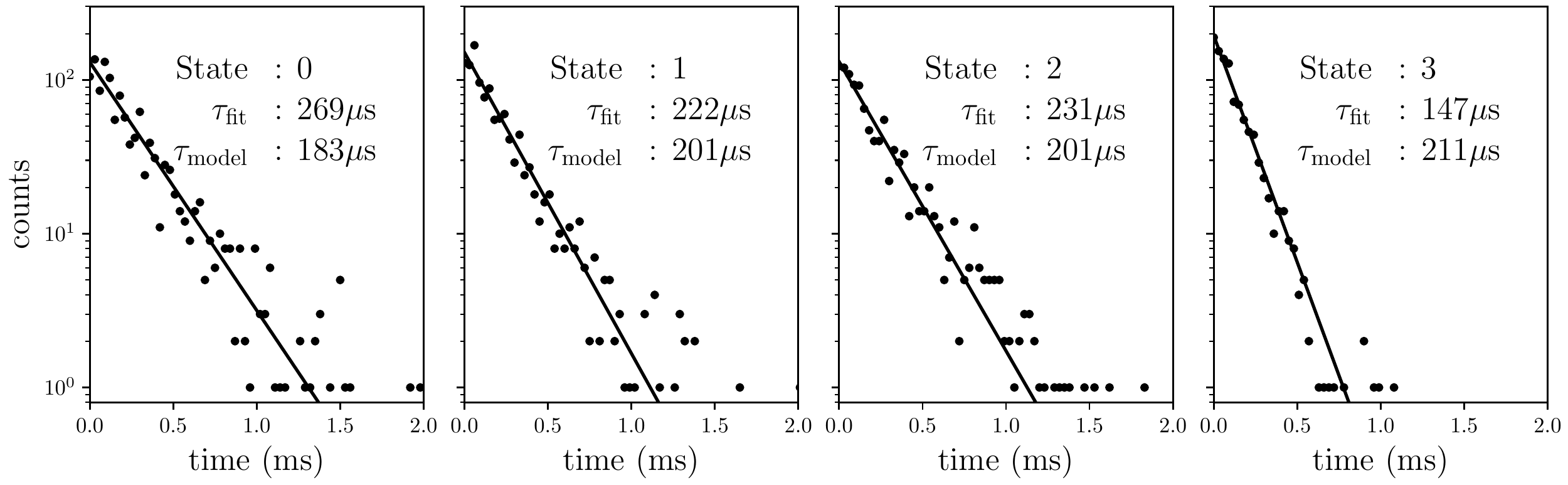}
    \caption{Histograms of the lifetimes of the four quasiparticle states (dots). The lifetimes are inferred from the jump times from the sequence predicted by the hidden Markov model. The fit is to an exponential (line), and the decay time is shown as $\tau_{\mathrm{fit}}$. The decay rate predicted by the HMM is also shown as $\tau_{\mathrm{model}}$.}
    \label{fig:jump_rates}
\end{figure*}

We further analyse the statistics of the quasiparticle tunneling events in \cref{fig:time-series}. In particular, based on the jump times predicted by the HMM, we estimate the lifetimes of the quasiparticle states, which are shown in \cref{fig:jump_rates}.


    
    

\clearpage


\end{document}